\documentclass{article}
\usepackage {graphicx}
\begin{document}
\baselineskip .75cm 
\begin{titlepage}
\title{\bf Landau's statistical mechanics for quasi-particle models}       
\author{Vishnu M. Bannur  \\
{\it Department of Physics}, \\  
{\it University of Calicut, Kerala-673 635, India.} }   
%\date{}
\maketitle
\begin{abstract}

Landau's formalism of statistical mechanics \cite{la.1} is applied to the quasi-particle model of quark gluon plasma. It is a general formalism and consistent with our earlier studies \cite{ba.1} based on Pathria's formalism \cite{pa.1}. Both the formalisms are consistent with thermodynamics and statistical mechanics. Under certain conditions, which are wrongly called thermodynamic consistent relation,  we recover other formalism of quasi-particle system \cite{go.1}, widely studied in quark gluon plasma.      
\end{abstract}
\vspace{1cm}
                                                                                
\end{titlepage}
%%%%%%%%%%%%%
\section{Introduction :}

Quasi-particle models of quark gluon plasma (qQGP) are phenomenological models to explain the non-ideal behavior of quark gluon plasma (QGP), seen in lattice simulation of quantum chromodynamics (QCD) at finite temperature \cite{bo.1,bo.1b,fo.1,pn.1} and relativistic heavy ion collisions. There are various quasi-particle models, which may be broadly classified into two groups depending upon two approaches. One approach, say model-I, was advocated first by Goloviznin and Satz \cite{gs.1,go.1,bu.1,ca.1,pe.1,bl.1} where the thermodynamics (TD) of quasi-particle system of quarks and gluons is developed by starting from the standard ideal gas expression for pressure, and all other TD quantities are obtained from it. The second approach, model-II 
\cite{ba.1,ba.3,ba.3b,ba.3c,ba.3d,sr.1,yi.1,zh.1,co.1,zu.1,ji.1}, starts from energy density,  which is a defined thermodynamic variable in ensemble theories, and then all other TD quantities are derived from it. In model-I, one needs certain condition for thermodynamic consistency which was called TD consistent relation. In model-II, it is not necessary because it is consistent with statistical mechanics and thermodynamics from the start \cite{ba.4}. Yet, under certain extra conditions it leads to model-I \cite{ba.1}. All these phenomenological models have few free parameters. There were also some attempts to unify above two models \cite{br.1,ga.1} in a thermodynamically self-consistent way.

Unfortunately, all models fit the lattice gauge theory (LGT) results by varying the free parameters of the model. Hence, one may not be able to differentiate the models based on the fits to LGT \cite{ba.4} data. Model-I has more than two parameters and model-II has only one.  

Of course, there are other phenomenological models like strongly coupled plasma models sQGP \cite{sh.1}, SCQGP \cite{ba.5}, etc. which were developed to study the non-ideal behaviour of QGP. One looks for these QCD motivated phenomenological models because of the limitations of perturbative QCD (pQCD). Even at high temperature like 1000 $T_c$, pQCD with expansion of the order of $g^6$ needed to fit the LGT data \cite{fo.1} and again with one fitting parameter. However, it fails to fit near $T_c$ up to $T \approx 3 T_c$. It is interesting to note that phenomenological models like qQGP and SCQGP with single parameter are able to fit LGT data of Wuppertal-Budafest group \cite{fo.1}  from 1.5 $T_c$ to 1000 $T_c$ \cite{ba.4,ba.6}, reasonably well. Near $T = T_c$ ( $T_c < 1.5 T_c$) models based on plasma may not be applicable and models like sQGP \cite{sh.1}, monopoles \cite{ch.1}, etc. may become relevent.  

Here we comment on the extensively used Landau's formalism of statistical mechanics for QGP and compare our results with qQGP models discussed above. We see that the standard statistical mechanics of Landau \cite{la.1} may be used to study QCD motivated quasi-particle models and leads to the modification of expressions for derivable thermodynamic quantities like pressure, entropy, etc. in contradiction to earlier works 
\cite{gs.1,go.1,bu.1,ca.1,pe.1,bl.1}. Only the expression for energy density has the ideal gas form, but pressure, entropy are all have extra temperature dependent term in addition to ideal gas expression due to temperature dependent quasi-particle mass. But, these results are consistent with the Pathria's formalism \cite{pa.1} of statistical mechanics of qQGP \cite{ba.1}.     

\section{Landau's formalism:} 

Many authors of qQGP (model-I) start from the expression for pressure, P, of ideal gas, which they claim to follow Landau's formalism. However, here we would like to point out that above is true only for particles with constant mass. Of course, it was pointed out and discussed in detail by Gorenstein and Yang \cite{go.1}, but their demand that the expressions for both pressure and energy density must be in the form of ideal gas, is over specification or constraints. In Pathria's formulation of statistical mechanics of qQGP \cite{ba.1}, energy density is defined as a statistical average in canonical or grand canonical ensemble and all thermodynamical quantities, including pressure, are derived from it. The energy density is in the form of ideal gas, but the pressure is not. Simultaneous demand that energy density and pressure must be of ideal gas form leads to thermodynamical inconsistency, requiring extra constraints to satisfy thermodynamic relations. If one wants to follow Landau's formalism, i.e., to develop statistical mechanics and thermodynamics, starting from pressure, we may follow Landau's formalism with external conditions \cite{la.1} (see page 109 to 110). In quasi-particle models, the thermal mass which is a function of temperature, may be taken as external parameter. 
That is, mass is externally controlled, depending on the temperature of reservoir in canonical or grand canonical ensembles. Following this concepts, we may  start with Gibb's distribution function,
\begin{equation}
w_n = e^{\alpha + \beta E_n} \,\, ,
\end{equation}
where $E_n(V,\lambda_1, \lambda_2, ...)$ which depends on external parameters $V, \lambda_1, \lambda_2, ...$ . Here, the external parameter may be volume $V$ and thermal mass $m_{th}(T)$. $\beta \equiv -1/T$, where $T$ is the temperature of the system which is equal to temperature of the reservoir in canonical ensemble. 

It must be stressed here that statistical mechanics is a probabilistic theory and average quantities are related to thermodynamic quantities. One can not twist and reformulate statistical mechanics for thermodynamics. For e.g., here, one asks the question what is the probability for the system to have energy, say, $E_r$. Because, the system in CE is in thermal contact with the reservoir and keeps on exchanging energy and hence $E_r$ fluctuates around the average value $U = \bar{E_r}$, which we call as the thermodynamic internal energy. Once we get $U$ all other thermodynamic quantities like pressure, entropy, etc. may be obtained from thermodynamic laws. No need of separate application of statistical mechanics for pressure and entropy.  
 
We consider here QGP with chemical potential $\mu =0$ and hence canonical ensemble formalism is sufficient.  
That is, we consider QGP made up of quasi-particles of quarks and gluons in thermal equilibrium with temperature $T$. In canonical ensemble $T, V, N$ are variables and hence as $T$ changes $m_{th}(T)$ changes and therefore $m_{th}(T)$ acts like an external parameter. $\alpha$ is the normalization factor such that 
\begin{equation}
\sum_n  e^{\alpha + \beta E_n} = 1\,\, , \label{eq:g}
\end{equation}
or,
\begin{equation}
\alpha = - \ln (\sum_n e^{\beta E_n}) = - \ln Q_N (T,V) \,\,,  \label{eq:al}
\end{equation}
In ref. \cite{la.1}, Landau formally chose $\alpha = \frac{F}{T}$, where $F$ is the Helmholtz free energy or thermodynamic potential. However, we, at this stage, treat $\alpha$ as a normalizing factor (Eq.(\ref{eq:al})), and compare with thermodynamic relations and fix it later.  
$Q_N$ is the canonical ensemble partition function. Following Landau \cite{la.1}, differentiating Eq.(\ref{eq:g}) with it's dependence, namely $\alpha, \beta, V$, etc. 
\begin{equation}
\sum_n  e^{\alpha + \beta E_n} \left(\delta \alpha +  E_n \delta \beta + \beta \frac{\partial E_n}{\partial \lambda_i} \delta \lambda_i \right) = 0\,\, , 
\end{equation}
where $\lambda_i$ may be external parameters like volume $V$, temperature, etc. Using the definition of average, 
\begin{equation}
\delta \alpha +  \bar{E} \delta \beta + \beta \overline{\frac{\partial E_n}{\partial \lambda_i}} \delta \lambda_i = 0 \,\, ,
\end{equation} 
where bar refers to statistical average. 
It may be further reduced into, 
\begin{equation}
\delta \bar{E} - \overline{ \frac{\partial E_n}{\partial V} } \delta V = \frac{\delta (\alpha + \beta \bar{E})}{\beta} + \overline{\frac{\partial E_n}{\partial m}} \frac{\partial m}{\partial \beta} \delta \beta \,\, , \label{eq:var}
\end{equation}
which may be compared with thermodynamic relation,
\begin{equation}
\delta U + P \delta V = T \delta S \,\, , \label{eq:td}
\end{equation}
and we may identify the entropy as,
\begin{equation}
S = -\alpha + \frac{\bar{E}}{T} + \int^T \frac{d\tau}{\tau} \overline{\frac{\partial E_n}{\partial m}} \frac{\partial m}{\partial \tau} \,\, , 
\end{equation}
and $U=\bar{E}$, $P = - \overline{ \frac{\partial E_n}{\partial V} }$.  
From U and S, we may obtain Helmholtz free energy $F = U - T S = - P V$ and hence pressure P. Note that all these derived quantities $S$, $F$ and $P$ have an extra temperature dependent term in addition to that of ideal system. Thus, even the standard notion of entropy is not valid for quasi-particle system with variable mass. These results are consistent with our earlier formalism \cite{ba.1} following Pathria \cite{pa.1} and we were able to fit lattice data using just one parameter 
\cite{ba.1,ba.3,ba.3b,ba.3c,ba.3d,sr.1,yi.1,zh.1,co.1,zu.1,ji.1}. 
    
\section{Landau's formalism for qQGP with bag pressure:} 

Using Landau's formalism of statistical mechanics with external conditions, we develop a consistent thermodynamics which reduces to our earlier qQGP which we developed following Pathria, where one starts from energy density and derive all other TD quantities. In this section, we redo the above calculations with temperature dependent bag pressure or zero point energy. In above section, we assumed that whole energy is used to excite  quasi-particles with thermal masses alone. Here, following Gorenstein and Yang \cite{go.1}, we introduce temperature dependent vacuum energy also. Hence, $E_n$ depends on another external parameter $B(T)$ in addition to the thermal mass $m(T)$ and  we have one more term due to variation in $B(T)$ which immediately leads to the modification of Eq.(\ref{eq:var}). Let us again start from Landau's steps with external parameters $m(T)$ and $B(T)$ ($\lambda_i$s) in addition to $V$. Hence, we have, 
\begin{equation}
\delta \alpha +  \bar{E} \delta \beta + \beta \overline{ \frac{\partial E_n}{\partial V} } \delta V + \beta \overline{ \frac{\partial E_n}{\partial \lambda_i} } \delta \lambda_i = 0 \,\, ,
\end{equation}
which may be reduced to 
\begin{equation}
\delta (T \alpha) =  -  \left(\frac{\bar{E}}{T} - \alpha \right) \delta T + \overline{ \frac{\partial E_n}{\partial V} } \delta V + \overline{ \frac{\partial E_n}{\partial \lambda_i} } \delta \lambda_i \,\, . \label{la.f}
\end{equation}
There is a summation over the index $i$. 
In Landau's case, where $V$ is the only external parameter and the last term in above equation is absent. Hence, by choosing $T \alpha = F$ and $\bar{E} = U$, above reduces to 
\begin{equation}
\delta F = - S \delta T + \overline{ \frac{\partial E_n}{\partial V} } \delta V \,\, ,
\end{equation} 
a valid thermodynamic relation. Note that $T \alpha = F$ is an assumption which leads to the thermodynamic relation  
\begin{equation}
\delta F = - S \delta T  - P \delta V \,\, .
\end{equation} 
In our case, here, $\lambda_i$ is $m(T)$ or $B(T)$ which are functions of $T$ and hence Eq.(\ref{la.f})  may be written as 
\begin{equation}
\delta (T \alpha) =  -  \left(\frac{\bar{E}}{T} - \alpha - \overline{ \frac{\partial E_n}{\partial \lambda_i} } \frac{\partial \lambda_i}{\partial T}\right) \delta T + \overline{ \frac{\partial E_n}{\partial V} } \delta V  \,\, .
\end{equation}
To compare above equation with valid thermodynamic relations, we have three possibilities. First one is the same as that of Landau, provided $\overline{ \frac{\partial E_n}{\partial \lambda_i} } \frac{\partial \lambda_i}{\partial T} = 0$ which immediately gives so called TD consistency relation and one arrives at model-I of quasi-particle models. Second choice may be to take internal energy $U = \bar{E} - T \overline{ \frac{\partial E_n}{\partial \lambda_i} } \frac{\partial \lambda_i}{\partial T}$ which however is inconsistent with the definition of average in statistical mechanics and hence not acceptable. Third choice may be, with little algebra, reduce it to the form     
\begin{equation}
\delta \left(T \alpha - T \int^T \frac{d\tau}{\tau} \overline{\frac{\partial E_n}{\partial \lambda_i}} \frac{\partial \lambda_i}{\partial \tau}\right) =  -  \left(\frac{\bar{E}}{T} - \alpha - \int^T \frac{d\tau}{\tau} \overline{\frac{\partial E_n}{\partial \lambda_i}} \frac{\partial \lambda_i}{\partial \tau}
\right) \delta T + \overline{ \frac{\partial E_n}{\partial V} } \delta V  \,\, .
\end{equation}
Now we chose $F = T \alpha - T \int^T \frac{d\tau}{\tau} \overline{\frac{\partial E_n}{\partial \lambda_i}} \frac{\partial \lambda_i}{\partial \tau} = - P V$ and $U = \bar{E}$ and above equation reduces to a valid thermodynamic equation Eq.(\ref{eq:td}) with $S = \frac{U}{T} - \alpha - \int^T \frac{d\tau}{\tau} \overline{\frac{\partial E_n}{\partial \lambda_i}} \frac{\partial \lambda_i}{\partial \tau}$.  Thus for pressure, for example, we get 
\begin{equation}
P =  - \frac{T}{V} \left( \alpha - \int^T \frac{d\tau}{\tau} \left[\overline{\frac{\partial E_n}{\partial m}} \frac{\partial m}{\partial \tau} +  \overline{\frac{\partial E_n}{\partial B}} \frac{\partial B}{\partial \tau} \right] \right) \,\, . 
\end{equation}
Substituting for $E_n$ which includes the vacuum energy contribution \cite{go.1} above reduces to 
\begin{equation}
P =  P_{id} - B(T) + \frac{T}{V} \int^T \frac{d\tau}{\tau} \left[\overline{\frac{\partial E_n}{\partial m}} \frac{\partial m}{\partial \tau} +  V \frac{\partial B}{\partial \tau} \right] \,\, . 
\end{equation}

Above expressions are thermodynamically consistent as it is. Both $m(T)$ and $B(T)$ are independent functions of $T$ which may be modelled. But, if we impose extra condition that 
\begin{equation}
\left[ \overline{\frac{\partial E_n}{\partial m}} \frac{\partial m}{\partial \tau} +  V \frac{\partial B}{\partial \tau} \right] = 0 \,\, , \label{eq:tcr}
\end{equation}
then we get qQGP model-I where $B(T)$ and $m(T)$ are related by above equation. Eq.(\ref{eq:tcr}) is what they wrongly called TD consistent relation, which is not necessary but simplifies the model. Our model-II corresponds to $B = 0$ which means that all energy is in quasi-particle modes. 
   
\section{Conclusions:} 

 Following the Landau's formalism  of statistical mechanics for a system subjected to external conditions \cite{la.1}, we developed the statistical mechanics  and thermodynamics of quasi-particle system of QGP. We arrive very naturally at our earlier formalism of qQGP \cite{ba.1}, developed using Pathria's formalism of  statistical mechanics \cite{pa.1}, where one starts from the expression for energy density. When we apply, the so called TD consistent relation, we get back the quasi-particle model-I \cite{go.1}. Therefore, the TD consistent relation is not needed to study QGP, and qQGP model with TD consistent relation may be a special case. If one starts from the ideal gas expression of pressure to develop thermodynamics of quasi-particle system, one always end up with thermodynamic inconsistency. Thus, Landau's formalism of statistical mechanics with external conditions \cite{la.1} clearly shows that the pressure for quasi-particle system is not in the form of ideal gas in contradiction with many quasi-particle models (model-I).

\end{document}